\documentclass[aps,prx,reprint,superscriptaddress,footinbib]{revtex4-2}

\usepackage[utf8]{inputenc}
\usepackage{xcolor}
\usepackage{appendix}
\usepackage{graphicx} 
\usepackage{epsfig}
\usepackage{epstopdf}
\usepackage{braket}
\usepackage[normalem]{ulem}
\usepackage[colorlinks]{hyperref}
\usepackage{bm}
\usepackage{siunitx}
\usepackage{amsmath}
\DeclareMathOperator{\Tr}{Tr}

\begin{document}

\title{Multimode ion-photon entanglement over 101 kilometers of optical fiber}

\author{V.~Krutyanskiy}
\email[Correspondence should be sent to ]{ viktor.krutianskii@uibk.ac.at}
\affiliation{Institut f\"ur Experimentalphysik, Universit\"at Innsbruck, Technikerstr. 25, 6020 Innsbruck, Austria}
\affiliation{Institut f\"ur Quantenoptik und Quanteninformation, Osterreichische Akademie der Wissenschaften, Technikerstr. 21a, 6020 Innsbruck, Austria}
\author{M.~Canteri}
\affiliation{Institut f\"ur Experimentalphysik, Universit\"at Innsbruck, Technikerstr. 25, 6020 Innsbruck, Austria}
\affiliation{Institut f\"ur Quantenoptik und Quanteninformation, Osterreichische Akademie der Wissenschaften, Technikerstr. 21a, 6020 Innsbruck, Austria}
\author{M.~Meraner}
\affiliation{Institut f\"ur Experimentalphysik, Universit\"at Innsbruck, Technikerstr. 25, 6020 Innsbruck, Austria}
\author{V.~Krcmarsky}
\affiliation{Institut f\"ur Quantenoptik und Quanteninformation, Osterreichische Akademie der Wissenschaften, Technikerstr. 21a, 6020 Innsbruck, Austria}
\affiliation{Institut f\"ur Experimentalphysik, Universit\"at Innsbruck, Technikerstr. 25, 6020 Innsbruck, Austria}

\author{B.~P.~Lanyon}
\affiliation{Institut f\"ur Experimentalphysik, Universit\"at Innsbruck, Technikerstr. 25, 6020 Innsbruck, Austria}
\affiliation{Institut f\"ur Quantenoptik und Quanteninformation, Osterreichische Akademie der Wissenschaften, Technikerstr. 21a, 6020 Innsbruck, Austria}

\date{\today}

\date{\today}

\begin{abstract}

A three-qubit quantum network node based on trapped atomic ions is presented. 
The ability to establish entanglement between each of the qubits in the node and a separate photon that has travelled over a \SI{101}{\kilo\meter}-long optical fiber is demonstrated. 
By sending those photons through the fiber in close succession, a remote entanglement rate is achieved that is greater than when using only a single qubit in the node. 
Once extended to more qubits, this multimode approach can be a useful technique to boost entanglement distribution rates in future long-distance quantum networks of light and matter.

\end{abstract}

\maketitle

 \section{Introduction}

Envisioned quantum networks consist of matter-based nodes for information processing and storage, that are interconnected with photonic links for the establishment of entanglement between the nodes \cite{Kimble2008, Wehnereaam9288}. Such networks could span distances from a few meters to a world-wide quantum network and would enable applications in computing, sensing and communication. 
Photon-mediated entanglement has been established across elementary networks consisting of two \cite{Moehring2007, Ritter12, Hofmann12, Bernien13, Delteil2016, Stockill2017, Magnard2020, Stephenson2020, Krut2022} and three \cite{Pompili2021} remote matter qubits, distributed over distances up to a 1.5 kilometers \cite{Hensen2015}.  
Recently, two atoms \SI{400}{\meter} apart were entangled over a spooled \SI{33}{\kilo\meter}-long fiber channel \cite{vanLeent2022}.

A key requirement for long-distance quantum networking is the ability to entangle a matter qubit with a photon and to distribute that photon over many tens of kilometers. That ability has been demonstrated using a range of different systems including trapped ions \cite{Bock2018, Walker2018, Krutyanskiy:2019cx} and atoms \cite{PhysRevLett.124.010510}, for distances of up to 50 kilometers \cite{Krutyanskiy:2019cx}.  
A second key requirement is the ability to integrate multiple quantum-logic capable qubits into network nodes \cite{Wehnereaam9288}. Nodes consisting of two co-trapped atoms  \cite{Langenfeld_Rempe_21}, two qubits in a diamond-defect system \cite{Kalb928} and two trapped ions \cite{Inlek2017, Hucul2014, PhysRevLett.130.213601, Drmota2022} have been demonstrated.

One advantage of having multiple qubits in network nodes is the possibility to perform \emph{multimode} entanglement distribution \cite{PhysRevLett.98.190503}. With a single matter-qubit, one has to wait at least the light travel time to learn if entanglement distribution was successful between nodes before trying again, or entanglement with the first photon is lost. 
For example, over 100 kilometers of optical fiber, the light travel time limits the maximum attempt rate for establishing remote entanglement to \SI{1}{\kilo\hertz}, which given the \SI{1}{\percent} transmission probability using standard optical fibers at  \SI{1550}{\nano\meter}, would yield a maximum possible success rate of  \SI{10}{\hertz}. 
This limit could be overcome by sending many photons into the channel, each entangled with a different matter-qubit in the node: thereby performing multiple entanglement distribution attempts within the single photon travel time.  

In this paper we present two main results. First, matter-photon entanglement is achieved over a spooled \SI{101}{\kilo\meter}-long fiber channel: twice the distance of previous works (e.g., \cite{Bock2018, Walker2018, Krutyanskiy:2019cx, PhysRevLett.124.010510}) and requiring a matter-qubit coherence time on the order of the photon travel time (\SI{500}{\micro\second}) to achieve.  Second, using three co-trapped matter qubits in the node we demonstrate a multimoding enhancement for the rate of entanglement distribution.

\section{Experimental setup and sequence}

\begin{figure*}[t]
	\begin{center}
        \includegraphics[width=2\columnwidth]{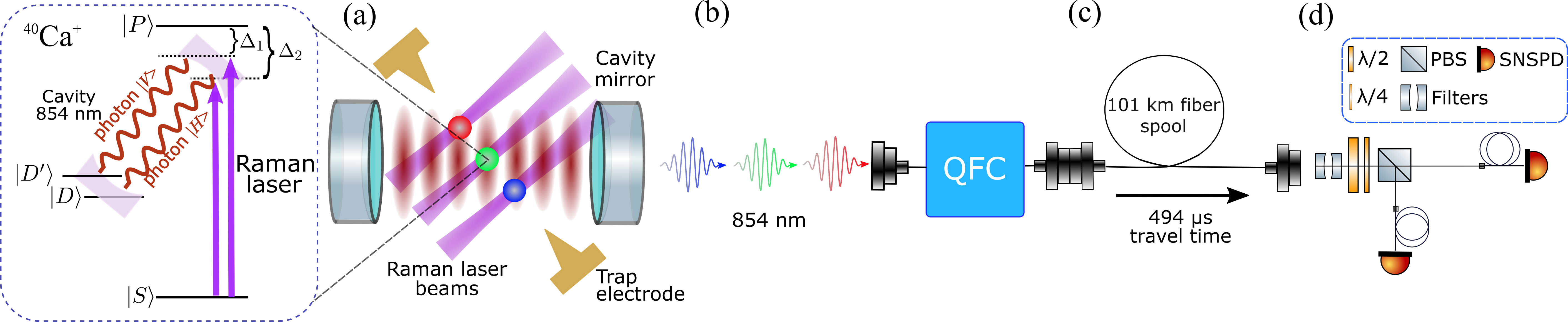}
		\caption{
		\textbf{Experimental schematic.} 
		(a) Three $^{40}$Ca$^+$ ions at neighbouring antinodes of an \SI{854}{\nano\meter} vacuum standing wave mode in an optical cavity. Sequential lasers pulses, one on each ion, generate three photons, each entangled by polarisation to the ion that emitted it. Inset: Atomic energy level diagram. $\ket{S}{=}\ket{4^{2} S_{1/2,m_j{=}{-}1/2}}$, $\ket{P}{=}\ket{4^{2} P_{3/2,m_j{=}-3/2}}$, $\ket{D}{=}\ket{3^{2}D_{5/2,m_j{=}-5/2}}$, $\ket{D'}{=}\ket{3^{2}D_{5/2,m_j{=}-3/2}}$. 
The frequency difference $\Delta_2-\Delta_1$ is equal to the one between $|D'\rangle$ and $|D \rangle$.
(b) The photons are converted to \SI{1550}{\nano\meter} via quantum frequency conversion (QFC) using the system of \cite{Krutyanskiy:2019cx, Krutyanskiy2017}. 
(c) A \SI{101}{\kilo\meter}-long single mode fiber spool (SMF-28). 
(d) Polarisation analysis involving half ($\lambda/2$) and quarter ($\lambda/4$) waveplates, filter network, a polarising beam splitter (PBS) and superconducting nanowire single photon detectors (SNSPDs). The narrowest element of the filter network is an air-spaced Fabry-P\'erot cavity with a \SI{250}{\mega\hertz} linewidth centred at \SI{1550}{\nano\meter} \cite{Krutyanskiy:2019cx}.}
    		\label{fig:concept}
	\end{center}
\end{figure*}

A conceptual schematic of the experimental setup is presented in Figure \ref{fig:concept} and now summarised.  
Our network node includes three $^{40}$Ca$^+$ ions confined in a 3D linear Paul trap and at the position of the waist of an optical cavity for photon collection at \SI{854}{nm} \cite{SchuppPRX2021, schuppthesis}. The ions are positioned at anti-nodes of the vacuum cavity standing wave. 
Because the cavity axis is not quite perpendicular to the ion-string axis (differing by a designed angle of 5 degrees), it is not possible to position the ions in the same anti-node. Instead, there is a unique axial confinement (and corresponding axial centre-of-mass frequency $\omega_z$) at which the ions can be positioned in neighbouring anti-nodes. 
A calibration process is performed to find that unique value, see Appendix \ref{positioning}, yielding $\omega_z=~$\SI{0.869(20)}{\mega\hertz}.  
Given $\omega_z$, we calculate that the ion spacing is \SI{5.26(10)}{\micro\meter} and that the angle between the ion string and the cavity axis is $85.3(1)^\circ$. 

Single photons are generated via a bichromatic cavity-mediated Raman transition (BCMRT) \cite{Stute2012, SchuppPRX2021}, driven via a  \SI{393}{\nano\meter} Raman laser beam with a \SI{1.2}{\mu }m waist at the ions \cite{PhysRevLett.130.213601}. 
A Raman laser pulse on an ion in the state $\ket{S}=\ket{4^{2} S_{1/2,m_j{=}{-}1/2}}$ ideally generates the maximally-entangled state 
\begin{equation}
\ket{\psi(\theta)}{=}(|D',V\rangle+e^{i\theta}|D,H\rangle)/\sqrt{2},
\label{eq_entnagl}
\end{equation}
where $|{D'}\rangle$ and $|{D}\rangle$ are the respective Zeeman states $|3^2{D}_{5/2}, m_j = -3/2\rangle$ and $|3^2{D}_{5/2}, m_j = -5/2\rangle$, $|{V}\rangle$ and $|{H}\rangle$ are the respective vertical and horizontal polarization components of a single photon emitted into the cavity vacuum mode, and $\theta$ is a phase set by the relative phase of the two freqeuency components in the bichromatic beam \cite{Stute2012}. After exiting the vacuum chamber through an optical viewport, photons are coupled into single mode optical fiber and then converted to \SI{1550}{nm} (telecom C band) via the polarisation-preserving single-photon frequency conversion system of \cite{Krutyanskiy:2019cx, Krutyanskiy2017}. 
Telecom photons are then sent into a 101km-long single mode fiber spool with a calculated photon travel time of \SI{494}{\micro\second} and measured total transmission probability of $1.36(4)\%$. 
Neither the optical length nor temperature of the fiber spool is actively stabilised. 
Finally, the photon polarisation is analysed in a chosen basis using a combination of motorised waveplates, a polarising beam splitter and two superconducting single photon nanowire detectors (Figure \ref{fig:concept}(d)).

The experimental pulse sequence consists of three parts: initialisation, photon generation and ion-qubit measurement. 
Initialisation consists of \SI{7}{\milli\second} of Doppler cooling followed by \SI{20}{\micro\second} of optical pumping into the state $\ket{S}$. 
Photon generation consists of a sequence of pulses, that we call an attempt, which is repeated up to 15 times (15 attempts). 
Each attempt begins with \SI{50}{\micro\second} of Doppler cooling and \SI{20}{\micro\second} of optical pumping, which serve to reinitialise the ions after any previous attempt. Next comes a \SI{50}{\micro\second} 
Raman laser pulse on each ion sequentially, spaced by \SI{12}{\micro\second} to allow e.g., the laser focus to be switched between ions using an acoustooptic deflector. The ideal result is a train of three photons, in which each photon is maximally entangled with the ion that emitted it. Next comes a \SI{503}
{\micro\second} wait time to allow all three photons to traverse the \SI{101}{\kilo\meter} fiber spool and be detected. 
At the beginning of that wait time, the $\ket{D}$ electron population of all ions is moved to the state $\ket{S}$ via an \SI{6.4}{\micro\second} $\pi$-pulse using a laser at \SI{729}{\nano\meter}. 
As such, the ion-qubits are encoded in superpositions of the $\ket{S}$ and $\ket{D'}$ states while the photons travel. 
After \SI{243.6}{\micro\second} of the wait time from the last 729-nm  pulse, a \SI{729}{nm} $\pi$-pulse then swaps the  $\ket{S}$ and $\ket{D'}$ population of all ion qubits, realising a spin echo.  
The pulse sequence for a single attempt is now completed. In the cases in which no photons are detected within the expected arrival time windows, another attempt is performed. 
In the cases in which at least one photon is detected within the expected arrival time windows, further attempts are aborted and ion-qubit measurement is executed. 

Ion-qubit measurement begins with an optional \SI{729}{\nano\meter} $\pi/2$-pulse implemented on the $\ket{S}$ to $\ket{D'}$ transition on all ions. The optional pulse is implemented when the ion-qubits are to be measured in the Pauli $\sigma_x$ or $\sigma_y$ basis: we set the optical phase of the pulse to determine in which of the two bases the measurement is made. The optional pulse is not implemented when the ion-qubit is to be measured in the $\sigma_z$ basis. 
Finally, single-ion resolved state detection is performed via electron shelving for \SI{1.5}{\milli\second} on all three ions simultaneously, at which point the experimental sequence is concluded. 
The chosen ion-qubit measurement basis and photon polarisation-qubit measurement are fixed throughout a single execution of the experimental pulse sequence. 

\twocolumngrid

The experimental pulse sequence is repeated sufficiently many times, and in sufficiently many measurement bases, to allow for reconstruction of the two-qubit states $\rho_{ij}$ of all nine possible combinations of one ion-qubit ($i$) and one photon-qubit ($j$), via state tomography.
States are reconstructed via the maximum-liklihood method and are conditional on successful detection of photon $j$.  Uncertainties in parameters derived from the states $\rho_{ij}$ are obtained via the Montecarlo technique. We use the concurrence $C$ \cite{Hill97_concurrence} to quantify the degree of entanglement in the states $\rho_{ij}$, where $0\leq C \leq 1$ and $C=1$ is a maximally entangled state achieved e.g., by the state of Equation 1.

\section{Experimental results}

In our first experiment we characterise the ion-photon states $\rho_{ij}$ before the photon conversion setup, at the emitted photon wavelength of  \SI{854}{\nano\meter}. For those states we use the new notation $\rho^0_{ij}$, reflecting that the photons have traveled over zero kilometers of fiber. A modified setup is used in which the fiber-coupled photons after the cavity output are sent to a polarisation analysis setup that is similar to the one shown in Figure \ref{fig:concept}, but with optics and detectors optimised for \SI{854}{\nano\meter}. The experimental pulse sequence has the following differences compared to the one described in the previous section: only one attempt to make a photon train is made per sequence, the \SI{503}{\micro\second} wait time is removed as well as the spin echo, and a Raman pulse length of \SI{60}{\micro\second} is used on each ion. 

 \begin{figure}[t]
	\begin{center}
        \includegraphics[width=1\columnwidth]{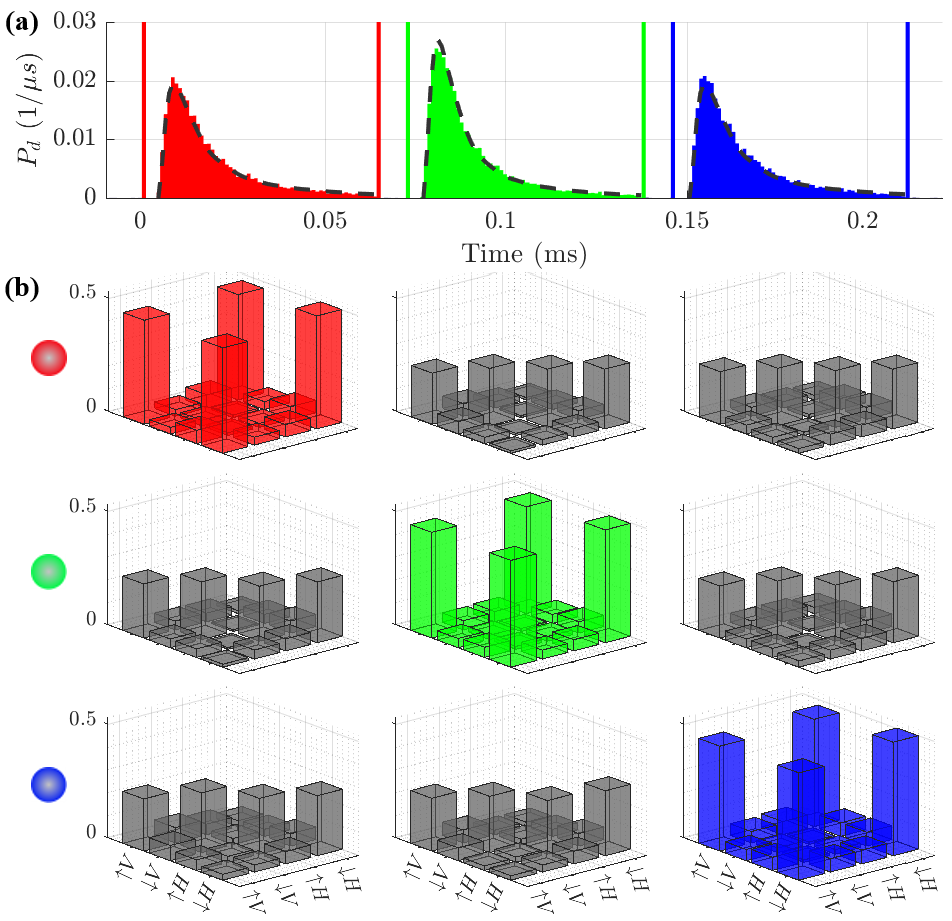}
		\caption{
		\textbf{Ion-photon entanglement over 0 km}. 
			{(a)} Histograms of \SI{854}{\nano\meter} photon arrival times. Probability densities are shown on the vertical axis: number of counts normalized by the number of attempts A and by the time-bin width of \SI{1}{\micro\second}, measured before QFC in Figure \ref{fig:concept} and without the fiber spool. Three single photon wavepackets are visible. Color is used to demark the ion that is expected to have produced the photon, following the ion colouring scheme in Figure \ref{fig:concept}. Time zero is when an acousto-optic modulator received a radio frequency signal to send a laser pulse to ion 1. Detection efficiencies and quantum states are determined within the \SI{65}{\micro\second}-long time windows shown via coloured vertical lines. Dashed black lines show results of a theoretical model.  
			{(b)} Absolute values of measured density matrices $\rho^{0}_{ij}$ of all nine ion-photon pairs. The ion (photon) involved in each row (column) is constant and indicated by the colored ball on the left (photon wavepacket above). States $\rho^0_{i=j}$ are colored red, green and blue. States $\rho^0_{i\neq j}$ are shown in grey. }
    		\label{fig:854}
		\vspace{-5mm}
	\end{center}
\end{figure}

The modified pulse sequence was implemented over a 42 minute period during which $A=41645$ attempts were made to generate a photon from each ion. 
Figure \ref{fig:854}(a) shows a histogram of the single photon detection events, in which three single photon wavepackets are clearly visible. Photons detected in the first, second and third \SI{65}{\micro\second}-long time windows are the ones expected to have been produced due to the corresponding Raman laser pulse applied to the first, second and third ion, respectively. The total number of counts recorded in those windows are {13127, 14465, 13326}, corresponding to estimated detection probabilities for 854 nm photons of {0.315(3), 0.347(3), and 0.320(3)} where uncertainties are based on Poissonian photon detection statistics. 
Only in 14 attempts was more than one detection event registered in the same time window, illustrating the single photon character of our source.
In $N_{single}=18337$ cases, exactly one photon detection event was registered in one of the window. 
In $N_{double}=9037$ cases, exactly two photon detection events were registered in different windows. 
In $N_{triple}=1485$ cases, exactly three detection events were registered in different windows.
The total probability to detect at least one photon within one photon generation attempt was $(N_{single}+N_{double}+N_{triple})/A= 0.693(4)$. The expectation value of the number of photons detected in an attempt was $(N_{single}+2\times N_{double}+3\times N_{triple})/A=0.981(5)$.

Each measured single photon wavepacket in Figure \ref{fig:854}(a) is well described by a theoretical model based on a master equation with model parameters for each wavepacket that differ only in the values used for the ion-cavity coupling strengths of the corresponding ion (see Appendix \ref{numerics}). The differences in those values are consistent with the effect of the Gaussian profile of the vacuum cavity mode across the ion string.
The simulations include an overall detection path efficiency of 0.518 for each of the photons 
which is  consistent with a value of 0.53(3) obtained from independent calibrations (see Appendix \ref{efficiency}).
 The detection path efficiency includes all losses encountered by a photon after emission into the cavity, including the finite probability of exiting the cavity into the output mode (independently measured to be 0.78(2) \cite{SchuppPRX2021}) all the way to the average \SI{854}{nm} detector efficiencies (independently measured to be 0.87(2) \cite{SchuppPRX2021}). 
 The simulations predict probabilities of  0.575, 0.664, 0.575 for emission of photons into the cavity from ions 1, 2 and 3, respectively. 
Lower detected photon efficiencies are achieved in this work, compared to  \cite{SchuppPRX2021}, largely due to the lack of ground state cooling and a sub-optimal Raman laser beam direction with respect to the principle magnetic field (quantisation) axis. Both issues could be corrected by reconfiguring the experimental setup in future. 

In all presented density matrices in this paper we use the following notion for ion-qubit states: $\ket{D'}=\ket{\uparrow}$ and $\ket{S}=\ket{\downarrow}$.
Figure \ref{fig:854}(b) presents the absolute values of all nine tomographically reconstructed two-qubit ion-photon density matrices $\rho^{0}_{ij}$. 
The concurrences of the three states $\rho^{0}_{11}$, $\rho^{0}_{22}$ and $\rho^{0}_{33}$ are {0.90(1), 0.91(1), 0.92(1)},
 respectively, proving strongly entangled states. The concurrences of the remaining six states $\rho^{0}_{i\neq j}$ are zero. 
The fidelities of the absolute values of the states $\rho^{0}_{11}$, $\rho^{0}_{22}$ and $\rho^{0}_{33}$, with $\ket{\psi(0)}$ (Equation \ref{eq_entnagl} for $\theta=0$) are 0.945(6), 0.950(5), 0.952(4), respectively. The fidelities of all the states $\rho^{0}_{i \neq j}$ with the maximally mixed two-qubit state are greater than 0.96 to within three  standard deviations of uncertainty. 
We use the fidelity defined as $\Tr{(\rho^{0}_{i=j}\ket{\psi(0)}\bra{\psi(0)})}$.  

The entangled states $\rho^0_{i=j}$ are locally rotated with respect to each other. Specifically, the phases of the large coherence terms ($\ket{\downarrow,H}\bra{\uparrow, V}$ and its complex conjugate) are $0.731(5)\pi, 0.632(5)\pi, 0.530(7)\pi$, for $\rho^{0}_{11},\rho^{0}_{22}$ and $\rho^{0}_{33}$, respectively. 
Those phases are consistent with a $\sigma_z$ rotation of the ion-qubit states as a function of time due to an incorrect setting of the frequency difference between the two fields in the Raman laser drive by \SI{689}{\hertz}. That frequency difference should ideally be equal to the one between the $\ket{D}$ and $\ket{D'}$ states (Figure \ref{fig:concept}(a)). 
The incorrect setting was due to a miscalibration in the transition frequencies and could be reduced to below the Hertz level by a more careful calibration using \SI{729}{\nano\meter} spectroscopy. Alternatively, such frequency offsets can be corrected by spin echos implemented on the ion-qubits during the photon travel time, as we demonstrate in the next experiment. %
The physical origins of the remaining imperfections in the entangled states are not yet known and identifying them will be the subject of future work. We conclude from analysis of the data in Figure \ref{fig:854} that an \SI{854}{\nano\meter} photon can be generated that is strongly entangled with any desired ion in the string.

\begin{figure}[t]
	\begin{center}
        \includegraphics[width=1\columnwidth]{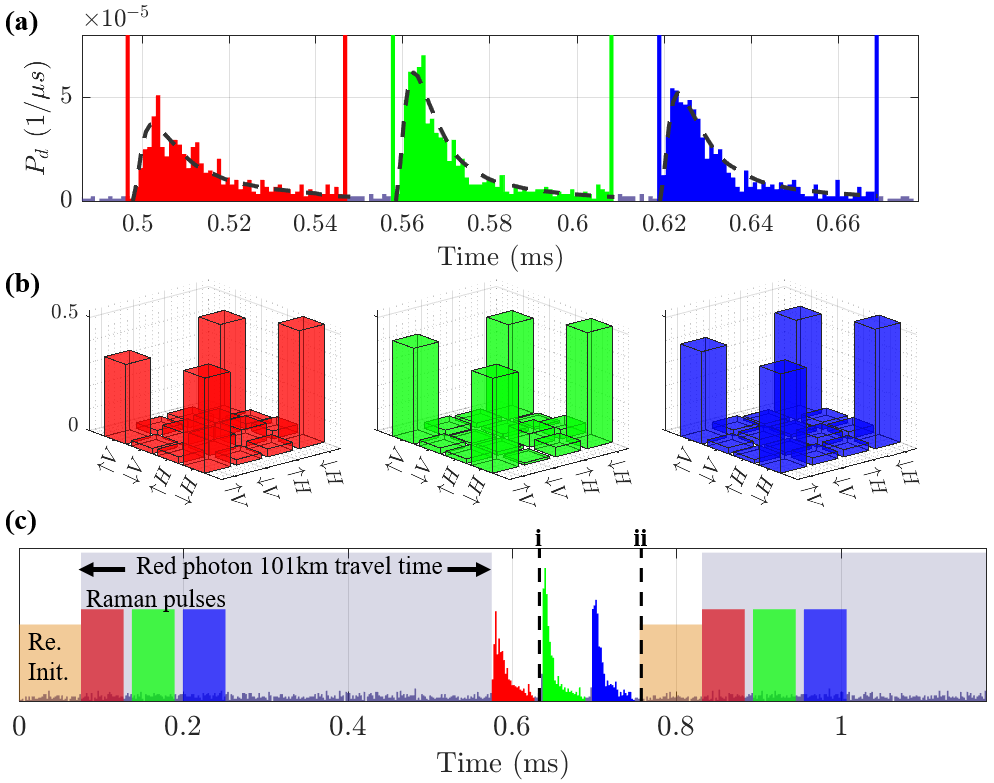}
		\caption{
		\textbf{Ion-photon entanglement over 101 km}. 		
{(a)} Histograms of \SI{1550}{\nano\meter} photon arrival times. Probability densities are shown on the vertical axis: normalized by the number of attempts A$^{100}$ and by the time-bin width of \SI{1}{\micro\second}, measured using the setup of Figure \ref{fig:concept}. The colour scheme is as described in Figure \ref{fig:854}. Dashed black lines show results of a theoretical model.  
{(b)} Absolute values of measured ion-photon density matrices $\rho^{100}_{i=j}$. The presented states are locally-rotated from the ones reconstructed directly from the data, as described in the main text. 
{(c)} Conceptual schematic of the experimental sequence. One attempt---involving three Raman laser pulses---took \SI{757}{\micro\second} (dashed line labelled ii). Attempts using a single ion would have taken \SI{633}{\micro\second} (dashed line labelled i). Re. Init. is the \SI{70}{\micro\second}-long cooling and optical pumping performed after each attempt. 
			}
    		\label{fig:100km}
		\vspace{-5mm}
	\end{center}
\end{figure}

In our second experiment the ion-photon states $\rho_{ij}$ are characterised using the full setup of Figure \ref{fig:concept}. For these states we use the new notation $\rho^{101}_{ij}$, reflecting that the photons have traveled over 101 kilometers of optical fiber. 
Measurements were taken over 45 minutes, during which $A^{101}=882,982$ attempts were made. 
Figure \ref{fig:100km}(a) shows a histogram of the recorded single photon detection events. 
The three photon wavepackets are spaced over a total of \SI{172}{\micro\second} and thus simultaneously fit well within the travel time of the fiber spool.
The total number of counts recorded in the three sequential \SI{50}{\micro\second}-long time windows were 572, 693 and 643, corresponding to detection probabilities of $p_1=6.5(3)\times 10^{-4}$, $p_2=7.8(3)\times 10^{-4}$, and $p_3=7.3(3)\times 10^{-4}$, respectively.
Only in two attempts was more than one detection event registered in the same time window. 
In $N_{single}=1900$ cases, exactly one photon detection event was registered in one of the windows. 
In $N_{double}=4$ cases, exactly two photon detection events were registered in different windows. 
There were no cases in which exactly three or more detection events were registered in different windows.
The total probability to detect (successfully distribute) at least one photon over 101km per attempt was $2.16(5)\times 10^{-3}$.
The expectation value of the number of photons detected in an attempt was also $2.16(5)\times 10^{-3}$.

The measured wavepackets of Figure \ref{fig:100km}(a) are well described by the ones from the master equation model. The only model parameter values that differ from those used to produce the simulations in Figure \ref{fig:854}(a) are: a lower total detection path efficiency of $1.26\times 10^{-3}$; the shorter Raman laser pulse used; a \SI{7}{\percent} lower Raman laser Rabi frequency and; ion-cavity coupling strengths that differ by up to \SI{10}{\percent}, consistent with an ion string shift of \SI{1.4}{\micro\meter} away from the cavity axis (see Appendix \ref{numerics}).  The lower path efficiency and Rabi frequencies are consistent with independent calibrations. The shift of the string was not calibrated, however, the calibration process used to position the middle ion in the centre of the cavity waist had not been performed for a month before the data was taken and therefore a relative drift of \SI{1.4}{\micro\meter} due to thermal effects is not unreasonable. 

Figure \ref{fig:100km}(b) presents the absolute values of the three tomographically-reconstructed states $\rho^{101}_{i=j}$, after the application of the same local two-qubit rotation was applied to each state. A local two-qubit rotation is a tensor product of single qubit rotations---one to the ion and one to the photon---which cannot change the entanglement content. 
The method used to obtain that local rotation is now described. First, the data from all matched ion-photon pairs ($i=j$) were added up and used to tomographically reconstructed a single ion-photon state $\rho^{101}$. Second, a numerical search was performed over local rotations that maximises the fidelity of $\rho^{101}$ with the state $\ket{\psi(0)}\bra{\psi(0)}$, yielding the optimum local rotation and a fidelity of 0.89(2). The concurrence of the state $\rho^{101}$ is 0.76(4). 
The concurrences of the states $\rho^{101}_{11}$, $\rho^{101}_{22}$ and $\rho^{101}_{33}$ are 0.71(8), 0.80(6), 0.83(6), respectively. After the local rotation, the fidelities of those states with $\ket{\psi(0)}\bra{\psi(0)}$ are 0.85(4), 0.88(3) and 0.90(3), respectively. 
No statistically significant rotation of the ion-qubits states with respect to each other is evident. In Appendix \ref{Infidelity} we describe a model of the effect of our photon detector background counts on ideal ion-photon entangled states. 
The results of the model show that the infidelities in the tomographically-reconstructed states $\rho^{101}_{i=j}$ are statistically consistent with the effects of those imperfections alone. Decoherence of the ion-qubits during the \SI{494}{\micro\second} photon travel time is insignificant: coherence times of \SI{62(3)}{\milli\second} are expected in our system when using optical spin echos \cite{PhysRevLett.130.213601}.

We turn now to consider the achieved multi-moding enhancement. 
Each attempt in the \SI{101}{\kilo\meter} experiment took $\tau= $ \SI{757}{\micro\second} (Figure \ref{fig:100km}(c)) and provided three opportunities to succeed in detecting a photon (one from each ion). The total probability for at least one successful photon detection per attempt was $P=$ 2.16(5)$\times10^{-3}$, which yields an effective success rate of $P/\tau=$ \SI{2.85(7)}{\hertz}. If instead we had used only one ion in the string, completing each attempt would have taken \SI{633} {\micro\second} (Figure \ref{fig:100km}(c)), as in addition to the reinitialisation and photon generation times,  one has to wait \SI{494}{\micro\second} for the photon to travel and be detected, before trying again. For the probability of success for that attempt we take the value from our experiment, for the most efficient central ion, of $p_2=7.8(3)\times 10^{-4}$. One then calculates a predicted effective success rate of \SI{1.23(5)}{\hertz} for the single-ion case. 
Therefore, by using all three ions we achieved a success rate increase by a factor of 2.3(1). That factor is reduced from the ideal value of three due to three separate effects: slightly lower photon emission probabilities from the ions not in the centre of the string; the times for switching the focus of the laser between the ions and; that we wait for all three photons to (potentially) arrive before trying to generate new photons. The last effect could be eliminated in future, after development of a single-ion-focused reinitialisation scheme.  Even considering a scenario in which the time for generating and reinitialising photons was effectively zero, such that each attempt took \SI{494}{\micro\second}, the 101km success rate for a single ion emitting in the string would still be only \SI{1.59(7)}{\hertz}.
.

\section{Conclusion and outlook}

In conclusion, ion-photon entanglement was achieved over a \SI{101}{\kilo\meter}-long fiber channel with a Bell state fidelity largely limited by detector background counts. The use of three co-trapped ion qubits allowed 
entanglement to be distributed at a higher rate than when using a single ion, by overcoming the attempt rate limit set by the photon travel time over the channel. 
In future, photon detection after the fiber channel could be used to swap entanglement to a duplicate remote ion-node via entanglement swapping \cite{Moehring2007,Ballance2020, Krut2022}. Here, each remote node sends a photon trains and coincident photon detection between different temporal pairs heralds entanglement of known remote ion pairs. 
The quantum processing and coherence times possible in ion-qubit registers could then be used to store the established entanglement for extended periods of time, as well as to purify the distributed entanglement and to grow the number of remote Bell pairs over time. 

The multimoding depth in our system could be significantly increased in future by coupling more ions in the node to travelling photons. For example, longer ion strings could perhaps be shuttled stepwise through the cavity mode by modulated the trap electrodes, allowing generation of a photon from each one. Alternatively, a single stationary ion could be used to generate photons sequential, and have its quantum state transferred to co-trapped ions after each attempt via quantum logic operations, as demonstrated for two ions in \cite{Drmota2022}. 
Benefiting from multimoding with hundreds or thousands of ions would require significantly shortening the current single photon wavepacket lengths (Figure \ref{fig:100km}(a)) such that they all fit simultaneously within the light travel time. Achieving that without compromising photon generation efficiency requires an increased ion-cavity coupling strength afforded e.g., by the smaller mode volume cavities \cite{Kobel2021,Teller:2023jx, christoforou2020}.  

Datasets are available online~\footnote{Will be provided upon acceptance}\\

\begin{acknowledgments}
This work was funded in part by the Austrian Science Fund (FWF) START prize  project Y 849-N20 and FWF Standalone project QMAP with project number P 34055.  This work received funding from 
the DIGITAL-2021-QCI-01 Digital European Program under Project number No 101091642 and project name `QCI-CAT', and
the European Union’s Horizon Europe research and innovation programme under grant agreement No. 101102140' and project name ‘QIA-Phase 1'. 
We acknowledge funding for V. Krutyanskiy by the Erwin Schr\"odinger Center for Quantum Science \& Technology (ESQ) Discovery Programme, and for B.P.L. by the CIFAR Quantum Information Science Program of Canada.
The opinions expressed in this document reflect only the author’s view and reflects in no way the European Commission’s opinions. The European Commission is not responsible for any use that may be made of the information it contains. For the purpose of open access, the author has applied a CC BY public copyright licence to any Author Accepted Manuscript version arising from this submission.

Experimental data taking was done by V.Kru., M.M., V.Krc. and M.C..
Development of the experimental setup was done by V.Kru., M.C., M.M, and B.P.L.. 
Data analysis and interpretation was done by V.Kru., M.C., M.M. and B.P.L.. 
Modelling was done by V.Kru.. 
The manuscript was written by B.P.L. and V. Kru., all authors provided detailed comments. 
The project was conceived and supervised by B.P.L..
\end{acknowledgments}

\begin{appendix}
\section{Positioning the ion string in the cavity}
\label{positioning}
The process is carried out in three steps. The goal of the first step is to overlap the centre of the ion trap (equivalently, the middle ion in the string) with the center of the waist of the cavity's \SI{854}{\nano\meter} $\mathrm{TEM_{00}}$ mode. 
This is achieved using a single ion following the method described in Appendix B. 1d of \cite{SchuppPRX2021}.

The goal of the second step is to obtain an ion-ion distance such that, when projected onto the cavity axis, the ions are spaced by \SI{427}{\nano\meter}: the distance between nodes (and anti-nodes) in the \SI{854}{\nano\meter} vacuum cavity standing wave. 
That is achieved by varying the axial confinement of the three-ion string and, for each value, performing measurements similar to the ones presented and explained in \cite{Beglev_2016}. Specifically, we record the \SI{854}{\nano\meter} photons when illuminating all three ions with a broadly-focused \SI{393}{\nano\meter} beam together with an \SI{854}{\nano\meter} and \SI{866}{\nano\meter} repumper. For each axial confinement we minimize the rate of the detected \SI{854}{\nano\meter} cavity photons by fine adjustment of the cavity position along its axis using an in-vacuum translation stage. The axial confinement that offers the lowest count-rate is found and interpreted as the situation in which each ion is located at a node of the vacuum-cavity standing wave. 

The goal of the third step is to position each of the three ions at a cavity anti-node. That is achieved using the single-ion focused Raman beam and repumpers to generate cavity photons from the central ion, then adjusting the cavity position along its axis using in-vacuum translation stages until the count rate is maximised.

\section{Numerical simulations of photon wavepackets}
\label{numerics}

Numerical simulations were performed to obtain an estimation for the photon generation efficiencies and single photon wavepackets. Specifically, the master-equation model of the laser-atom-cavity system is used from \cite{SchuppPRX2021}.  The model parameters include the experimental geometries of the Raman laser, cavity and magnetic field, which are the same as described in \cite{PhysRevLett.130.213601} (in particular, see Sec. I.B and III.C of the supplementary material).

A key model parameter is the maximum strength of the coherent coupling between a single photon in the cavity and a single ion, which is calculated to be $g_0 = 2\pi \times \SI{1.53}{\mega\hertz}$ in our system, based on the cavity geometry and the properties of the atomic transition. Here we consider the $\ket{P}-\ket{D}$ and $\ket{P}-\ket{D'}$ transitions but do not take into account the different Clebsch-Gordon coefficients for the two transitions or the projection of the transition polarizations onto the cavity-photon polarizations, both of which are accounted for separately in simulations~\cite{Stute2012}. 
The coupling strength of the bichromatic cavity-mediated Raman transition on a given ion in a string is reduced by e.g., the ion's motion in the trap and by any displacement of the ion's position away from the cavity axis. 
We model those effects using a reduced ion-cavity coupling strength for ion $i$ as $g_i = x_i \gamma g_0$, where $0\leq x_i \leq 1$ and $0\leq \gamma \leq 1$. 
The parameter $x_i$ quantifies any reduction in ion-cavity coupling strength due to ion $i$ not being positioned at the cavity axis. 
The parameter $\gamma$ quantifies any other reduction in the coupling strength of the bichromatic cavity-mediated Raman process  e.g., due to the motion of the ion in the trap. We use a single value for $\gamma$ for all ions and determine its value by comparing measured single-photon temporal wavepackets with simulated wavepackets based on numerical integration of the master equation for a range of values of the coupling strength~\cite{SchuppPRX2021}.\\

Another key model parameter is the strength of the bichromatic drive. 
In order to determine that strength we measure the AC Stark shift of the Raman transition via spectroscopy, as described in Sec. I.B and III.C of the supplementary material of \cite{PhysRevLett.130.213601}.
The bichromatic drive field polarization in the experiment (and simulations) is set to linear and perpendicular to the magnetic field and thus consists of  an equal superposition of two circularly polarized components $\sigma^-$ and $\sigma^+$. While only the $\sigma^-$ component is set to resonantly drive the bichromatic cavity-mediated Raman transition, both polarisation components contribute to the AC Stark shift. 
In the simulations we set the strength of the bichromatic drive $\Omega^-$---the Rabi frequency  with which the $\sigma^-$ transition $|S\rangle=|4^2S_{1/2}, m_j = -1/2\rangle$ to $|4^2P_{3/2}, m_j = -3/2\rangle$  is driven---to the value for which the model predicts the same total AC Stark shift as measured in the experiment. The model requires specifying both Rabi frequencies, $\Omega^-_1$, $\Omega^-_2$ of the two $\sigma^-$-polarized frequency components of the bichromatic drive. Here, $\Omega^-_1$ stands for the component that drives $\ket{S}-\ket{D'}$ and results in a vertically polarized ($V$) photon and  $\Omega^-_2$ stands for the component that drives $\ket{S}-\ket{D}$ and results in a horizontally-polarized ($H$) photon. 
For all the simulations we set $(\Omega^-_1)^2+(\Omega^-_2)^2 = (\Omega^-)^2$ and $\Omega^-_1/\Omega^-_2 = 0.81$: the value for which the model predicts equal probabilities for the generation of the $H$ and $V$ polarized photons.

Now we provide more information about the simulations for the experiment in which 854-nm photons are detected, shown in Figure \ref{fig:854} of the manuscript. 
By considering the ion string to be centred around the cavity axis and, from the Gaussian cavity mode profile, we calculate the values of $\{x_i\}$ to be $\{0.83, 1, 0.83\}$ for the three ions. 
Next, $\gamma = 0.784$ is found to provide a close match between the measured and simulated wavepackets. We measured an AC Stark shift of the Raman transition of \SI{0.88(2)}{\mega\hertz} for all the three ions. In simulations we use the value of $\Omega^- = 2\pi\times\SI{31.47}{\mega\hertz}$ for which the model predicts an AC Stark shift of the Raman transition of \SI{0.88}{\mega\hertz}. 

We now provide more information about the simulations for the experiment in which 1550-nm photons are detected, shown in Figure \ref{fig:100km} of the manuscript. We use $\gamma = 0.784$  as found using the data of Figure \ref{fig:854}. We use $\{x_i\}$ values of $\{0.739, 0.987 0.894\}$ which are calculated from the Gaussian cavity mode profile in the case of a \SI{1.4}{\micro\meter} displacement of the ion string along the trap axis direction with respect to the center of the cavity mode. This shift is found by comparing simulated wavepackets and another experiment performed on the same date as the one presented in Figure \ref{fig:100km} but involving \SI{50}{\kilo\meter} of fiber instead of \SI{101}{\kilo\meter} (in which the measurement statistics is better due to the higher photon detection efficiency). In the \SI{101}{\kilo\meter} experiment, we measured AC Stark shifts of the Raman transition of \SI{0.82(2)}{\mega\hertz} for all the three ions. In simulations we used the value of $\Omega^- = 2\pi\times\SI{30.41}{\mega\hertz}$ for which the model predicts AC Stark shifts of the Raman transition of \SI{0.82}{\mega\hertz}.

\section{Photon path efficiency}
\label{efficiency}
Here, a detailed efficiency budget is presented for the photon detection path. When not given explicitly, uncertainties in given probabilities are half of the last significant digit. The beginning of each element in the following list gives the probability associated with a distinct part or process in the experiment. The detection path efficiencies provided in the main text should be compared with a product of these probabilities (or a subset thereof, for the data taken involving \SI{854}{\nano\meter} photons). For the photons detected at \SI{854}{\nano\meter}, the total probability of the list is 0.53(3). For the photons detected at \SI{1550}{\nano\meter}, the total probability of the list is $15(1.2)10^{-4}$. 

\begin{enumerate}
\item 0.78(2): probability that, once a cavity photon is emitted into the cavity, the photon exits the cavity into freespace on the other side of the output mirror \cite{SchuppPRX2021}.   
\item 0.96(1): transmission of free-space optical elements that are between the cavity output mirror and first fiber coupler (see $P_{el}$ in \cite{SchuppPRX2021}).
\item 0.81(3): efficiency of coupling the photons into the first single mode fiber \cite{SchuppPRX2021}. This value should rather be considered an upper bound, since it was measured some days before the data presented in this paper was taken. We anticipate that coupling could be improved in future with better couplers and an anti-reflection coated fiber end facet. 
\item In case of detection of \SI{854}{\nano\meter} photons (Figure \ref{fig:854}) the lists ends here with 0.87(2): detection efficiency of either of the single photon detectors for \SI{854}{\nano\meter} \cite{SchuppPRX2021}. In case of detection of \SI{1550}{\nano\meter} photons (Figure \ref{fig:100km}) this item is not relevant and the list continues. \\

\item  0.30(1): 
fiber-input to free-space-output efficiency of the quantum frequency conversion setup together with the spectral filtering and polarization analysis optics (see panels (b) and (d) of Figure \ref{fig:concept}). This value was measured with laser light directly before acquiring the data presented in Figure \ref{fig:100km}. \\

\item 0.0136(4): measured transmissions of the 101-km fiber, consisting of two 50.4-km SMF-28 fiber spools and one fiber connector. See panel (c) of Figure \ref{fig:concept} for the position of the fiber in the optical path.\\

\item 0.95: transmission of a fiber joiner present in the path.\\

\item 0.83(3): efficiencies of coupling the telecom photons into the detector's single mode fibers, see panel (d) of Figure \ref{fig:concept}. 

 \item  0.75(2): detection efficiency of either of the telecom single photon detectors \cite{PhysRevLett.130.213601}.
\end{enumerate}

\section{Infidelity due to photon detector background counts}
\label{Infidelity}

We model the effect of background photon detector counts on the 1550-nm ion-photon states presented in Figure \ref{fig:100km} (defined as a detector click that didn’t result from a photon from the ion). 
For this, the background count rate is extracted from the measured counts in the tomography experiments by looking far outside the time windows in which the photons from the ions arrive and summing the contributions from both detectors, giving 2.0(1)-cps. 
The infidelity that those background counts would contribute, when added to a perfect maximally-entangled Bell state, is simulated numerically. Specifically, the expected background count probabilities in our photon time windows are added to the expected measurement outcome probabilities for a perfect state, then, after renormalisation, a new ‘noisy’ state density matrix is reconstructed via Maximum Likelihood tomography.
Using the background count rate of 2-cps we thereby calculate the maximum observable fidelities to be 0.88, 0.90, 0.90 for the three ion-photon pairs respectively, as ordered elsewhere in the manuscript. These fidelities can be compared with the ones obtained in the experiment of  0.85(4), 0.88(3) and 0.90(3).

\end{appendix}


%

\end{document}